\documentclass[11pt,a4paper]{book}
\usepackage{amsmath}
\usepackage{graphicx}
\usepackage{makeidx}
\usepackage{caption} 
\usepackage{paralist} 
\usepackage{fancyheadings}
\usepackage[%
  colorlinks=true,%
  urlcolor=blue,%
  citecolor=blue%
]{hyperref}
\usepackage[font={small,sl}]{caption} 
\usepackage[authoryear]{natbib} 

\def\cite{\citep}

\newcommand{\figref}[1]{\hyperref[fig-#1]{Figure \ref*{fig-#1}}}
\newcommand{\secref}[1]{\hyperref[sec-#1]{Section \ref*{sec-#1}}}

\newcommand{\remark}[1]{} 
\newcommand{\foothref}[2]{\href{#1}{#2}\protect\footnote{#2: \url{#1}}}

\makeatletter \@beginparpenalty=5000 \makeatother

\newcommand{\AF}{K.\@~Anton~Feenstra}
\newcommand{\SA}{Sanne~Abeln}
\newcommand{\JH}{Jaap~Heringa}

\makeindex

\begin{document}

\frontmatter

\pagestyle{fancy}
\lhead[\small\thepage]{\small\sf\nouppercase\rightmark}
\rhead[\small\sf\nouppercase\leftmark]{\small\thepage}
\newcommand{\innerfoot}{\footnotesize{\sf{\copyright} Abeln \& Feenstra}, 2014-2017}
\newcommand{\outerfoot}{\footnotesize \sf Structural Bioinformatics}
\lfoot[\outerfoot]{\innerfoot}
\cfoot{}
\rfoot[\innerfoot]{\outerfoot}
\renewcommand{\footrulewidth}{\headrulewidth}

\title{Structural Bioinformatics}
\author{\AF
  \and \SA
  \and
  \\[10ex]
  \textrm{\footnotesize Centre for Integrative Bioinformatics (IBIVU), and }\\
  \textrm{\footnotesize Department of Computer Science, }\\
  \textrm{\footnotesize Vrije Universiteit, De Boelelaan 1081A, 1081 HV Amsterdam, Netherlands}
}

\maketitle

\section*{Abstract}
This chapter gives a graceful introduction to problem of protein three-dimensional structure prediction, and focuses on how to make structural sense out of a single input sequence with unknown structure, the `query' or `target' sequence. We give an overview of the different classes of modelling techniques, notably template-based and template free. We also discuss the way in which structural predictions are validated within the global community, and elaborate on the extent to which predicted structures may be trusted and used in practice. Finally we discuss whether the concept of a single fold pertaining to a protein structure is sustainable given recent insights. In short, we conclude that the general protein three-dimensional structure prediction problem remains unsolved, especially if we desire quantitative predictions. However, if a homologous structural template is available in the PDB model or reasonable to high accuracy may be generated.  
\newpage

\tableofcontents

\mainmatter

\newcommand{\and}{\quad}

\newpage
\setcounter{chapter}{6}

\chapterauthor{\SA \and \JH \and \AF\\[5ex]
\textrm{\footnotesize Centre for Integrative Bioinformatics (IBIVU) and \\
Department of Computer Science, \\
Vrije Universiteit, De Boelelaan 1081A, 1081 HV Amsterdam, Netherlands}
}
\chapter{Introduction to Protein Structure Prediction}

\section{What is the protein structure prediction problem?}

\subsection{Predicting the structure for a protein sequence}
\begin{figure}[bht]
\includegraphics[width=\linewidth]{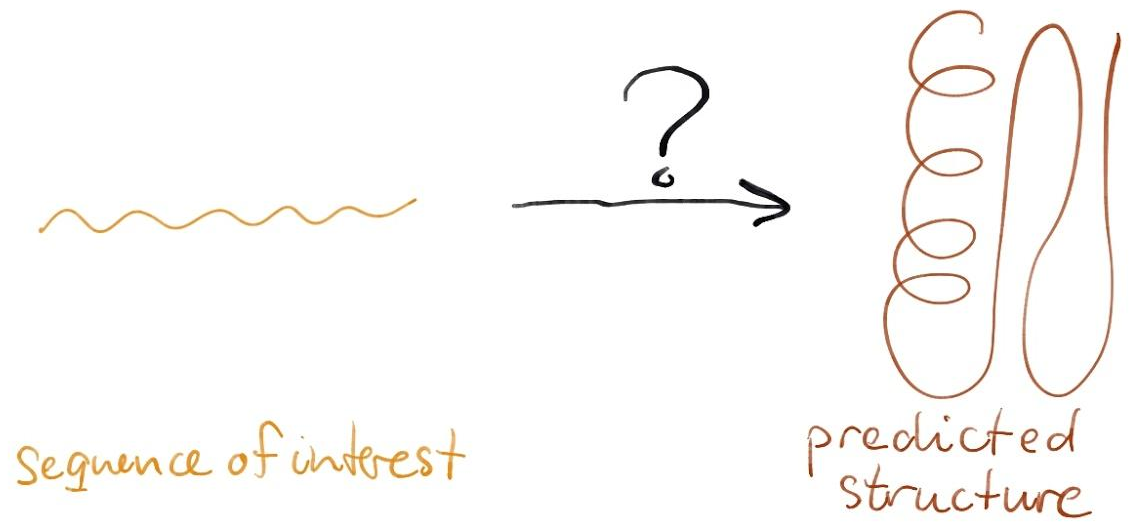}
\caption{Structure prediction methods try to answer the question: given an amino acid sequence, what is the folded protein structure?}
\label{fig-3DPred-Seq2Struc} 
\end{figure}

This chapter revolves around a simple question: ``given an amino acid sequence,  what is the folded structure of the protein?'' (\figref{3DPred-Seq2Struc}) Even though this seems like a simple question, the answer is far from straightforward. In fact, whether we can give an answer at all depends heavily on the sequence in question and available protein structures that can be used as modelling templates. While the number of structures deposited in the Protein Data Bank (PDB) \cite{Berman2000} continues to rise rapidly \footnote{\url{https://www.rcsb.org/pdb/statistics/contentGrowthChart.do?content=total\&seqid=100}}, the number of sequenced genes rises much faster. The large and widening gap between protein structures and sequences makes structure prediction an important problem to solve. Fortunately, recently developed methods can use these large resources of sequence data  to increase the quality of some predictions.  Here, we will give an overview of current structure prediction methods, and  describe some tools that provide insight into how reliable the structure predicted will be.

The typical problem is that we want to generate a structural model for a protein with a sequence, but without an experimentally determined structure. 
In this chapter, we will build up a workflow for tackling this problem, starting from the easy options that, if applicable, are likely to generate a good structural model, and gradually working up to the more hypothetical options whose results are much more uncertain. 

Another very important remark is in place here: the modelling strategy should depend heavily on what we want to do with the structure. Do we want to predict where the functional site of the protein is, whether a specific substrate binds, or if a certain residue may be exposed to the surface? These different questions imply a different degree of accuracy in the answer, and may lead to choices regarding technology and methods to carry out these predictions. It is important to keep in mind that one of the most important aspects of any scientific model is whether a research question may be answered with the model produced or not. Even if we do have an experimental structure available, some of these questions may not be straightforward to answer; we will come back to this issue later in the chapter.

\begin{figure}
\centerline{
  \includegraphics[height=0.8\linewidth]{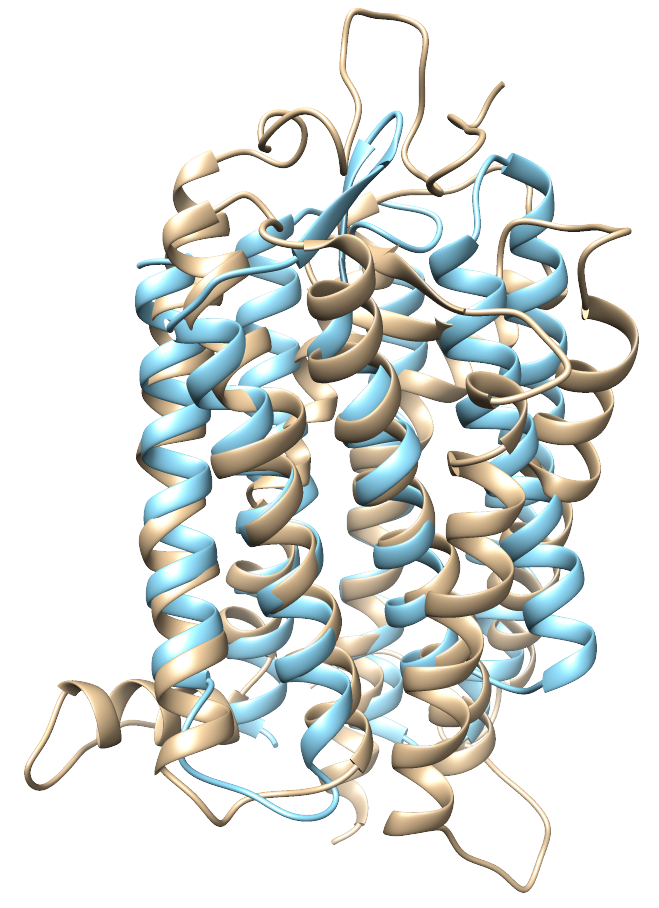}
  \includegraphics[height=0.8\linewidth]{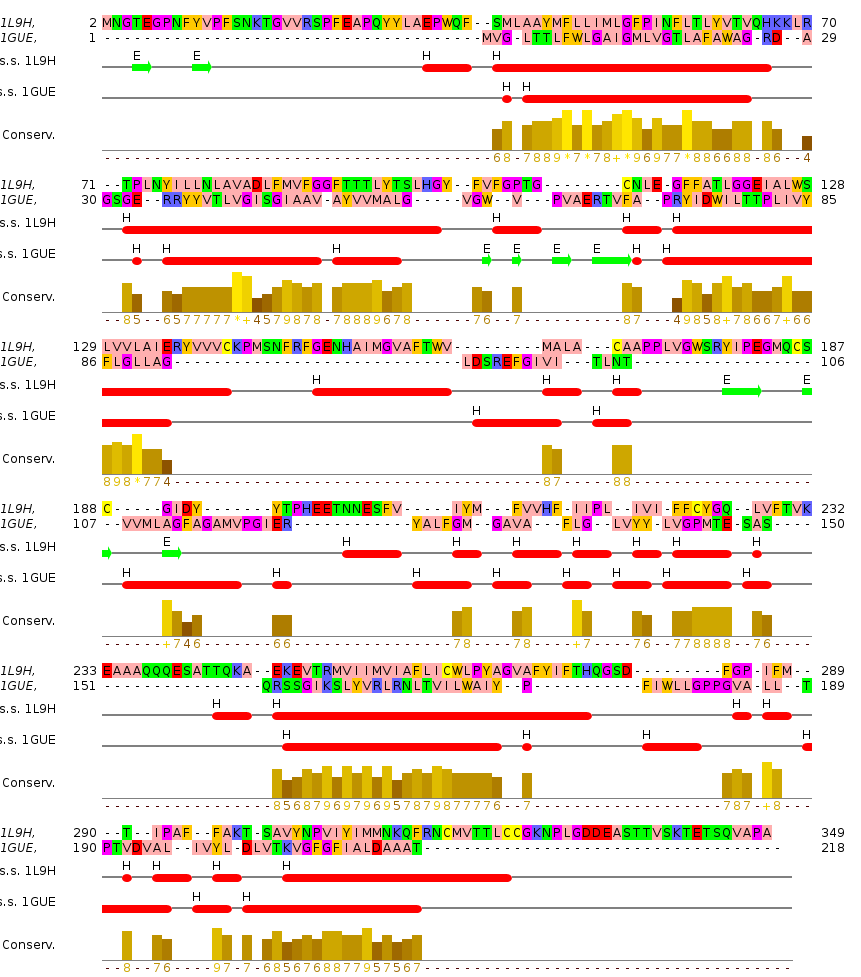}
}
\caption{Protein structure more conserved than sequence. Here the output of a structural alignment is shown on the left, created using \textsl{Chimera} \protect\footnotemark~\protect\cite{Pettersen2004}. The structural alignment shows both proteins are highly similar; the RMSD is 2.3 Å over 144 aligned residues. Furthermore, the function of the two proteins, one from cattle (\href{http://www.rcsb.org/pdb/explore.do?structureId=1l9h}{PDB:1L9H}, light brown) and one from a archaeon (\href{http://www.rcsb.org/pdb/explore.do?structureId=1gue}{PDB:1GUE}, light blue), is similar: both are light sensitive rhodopsins, used for vision and phototaxis, respectively. However, as can be seen in the sequence alignment on the right, the sequence identity is only 7\%. This is lower than would be expected for any two random sequences. The alignment shown is based on the structural alignment on the left, and visualised using \textsl{JalView} \protect\cite{Waterhouse2009}.}
\label{fig-3DPred-StrucCons} 
\end{figure}

\subsection{Structure is more conserved than sequence}

Almost all structure prediction relies on the fact that, for two homologous proteins, structure is more conserved than sequence (see \figref{3DPred-StrucCons})\footnotetext{Molecular graphics and analyses were performed with the UCSF Chimera package. Chimera is developed by the Resource for Biocomputing, Visualization, and Informatics at the University of California, San Francisco (supported by NIGMS P41-GM103311).}. 
The real power of this observation manifests itself when we turn this statement around: if two protein sequences are similar, these two proteins are likely to have a very similar structure. The latter statement has very important consequences. It means that if our sequence of interest is similar to a protein sequence with a known structure, we have a good starting point for a structural model. In such a scenario we use sequence similarity, suggesting an homologous relation between the proteins, to predict the structure. 
The vast majority of accurate structure prediction methods use structure conservation as an underlying principle; while methods that have been developed to deal with the more difficult modelling questions, exploit the sequence-structure-conservation relation in an advanced manner, as discussed towards the end of this chapter.

\begin{figure}
\centerline{\includegraphics[width=0.7\linewidth]{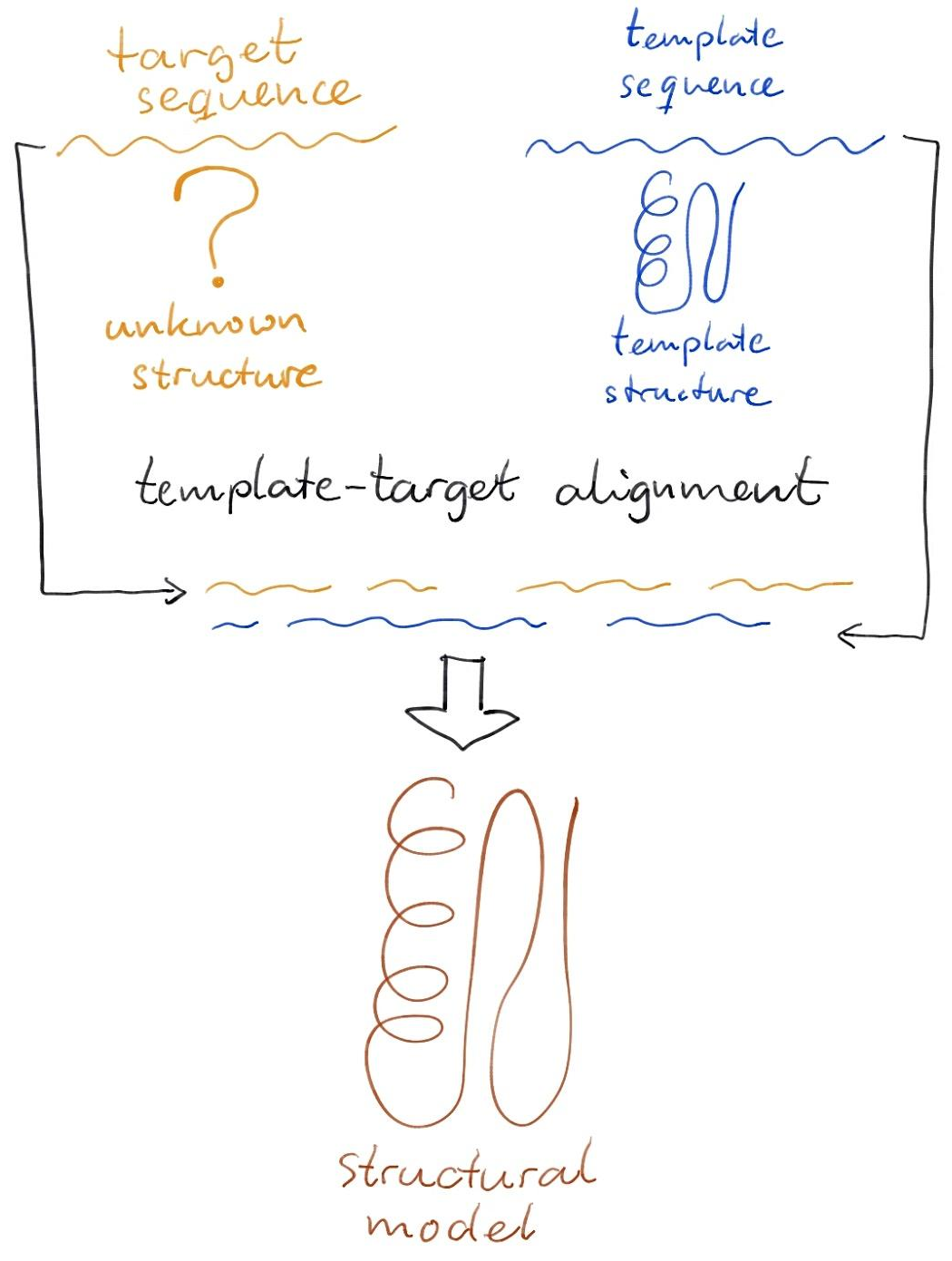}}
\caption{Terminology used in protein structure prediction. We start from our protein of interest (with no known structure): the target sequence. First step is find a matching protein: a template sequence with known structure; the template structure. We then create a template-target sequence alignment, and from this alignment create the structural model which is the solution structure for our target protein.}
\label{fig-3DPred-Terminology} 
\remark{Copyright OK: from scratch by Anton}
\end{figure}

\subsection{Terminology in structure prediction}

Firstly, we should take care to lay down a good problem definition. Here we will generously borrow the nomenclature from the Critical Assessment of Protein Structures (CASP). CASP is a scientific competition, in which structure prediction groups and structure prediction servers compete to predict the structure for an unknown sequence, that has been running since 1995 \cite{Moult1995}. The sequence for which we will predict a structure is called the \emph{target} sequence. If there is a suitable structure to build a model for our query sequence we call this structure a \emph{template}, see also \figref{3DPred-Terminology}. Using the structure of the template and using the \emph{sequence alignment} between the template and the target sequence, we can create one or more structural \emph{models}: the  predicted structure, for a target sequence. In CASP structural models from different prediction methods, are compared to the experimentally determined solution or target structure.

\begin{figure}
\centerline{\includegraphics[width=1.5\linewidth]{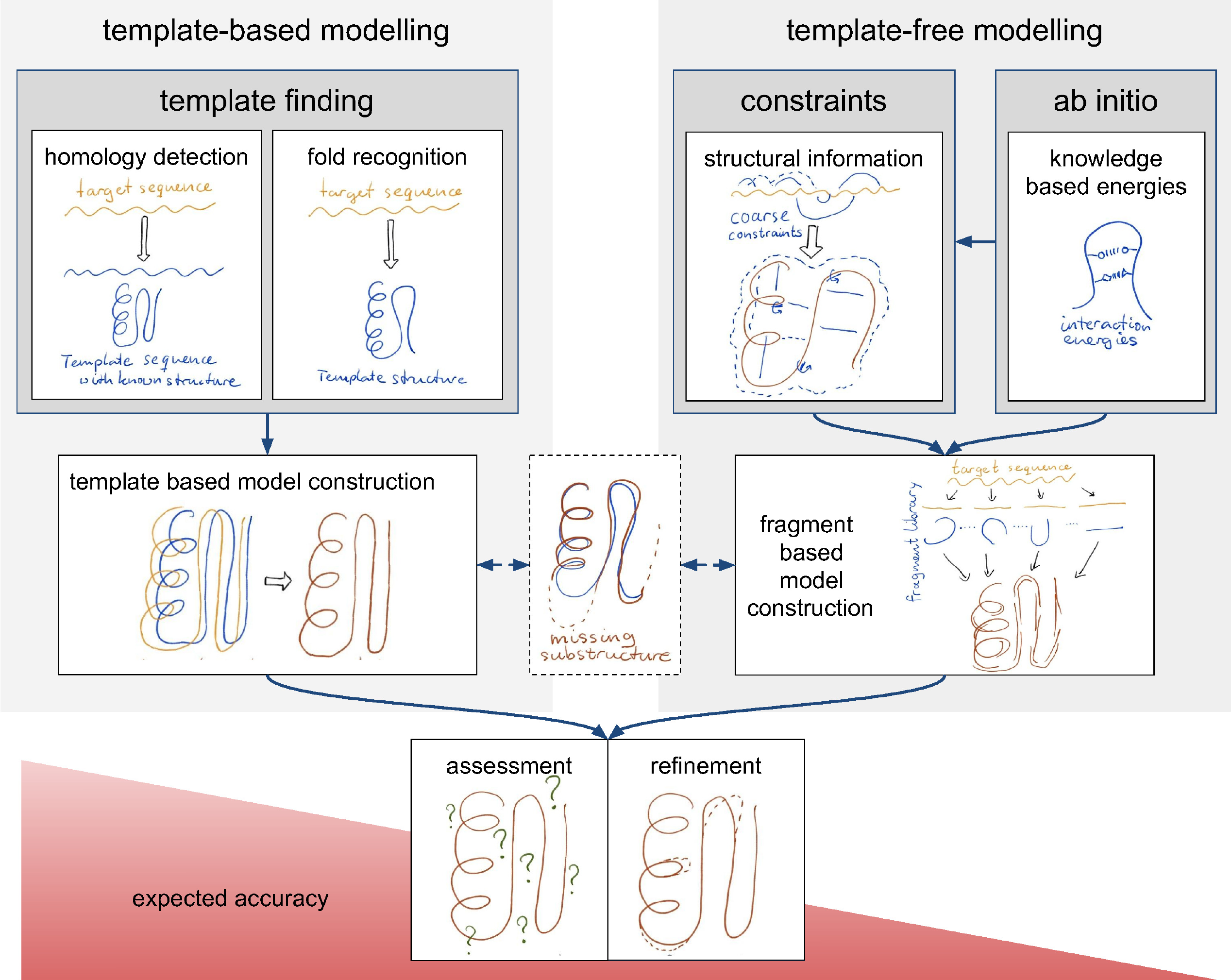}}
\caption{Overview of Structure Prediction. Template-based modelling: a template is found on the basis of homology between the template and the target. Fold recognition: no obvious homologous structure can be found in the PDB, we need fold recognition methods to find a suitable template. Template-free modelling: no suitable template for protein domains can be found. Without template, we need to use a combination of coarse constraints from experiment or co-evolution analysis, and ab initio prediction. Ab initio methods typically work with taking fragment templates from various proteins, and assemble these into a model or decoy structure. Expected model accuracy declines from left to right: good accuracy is expected if based on homology; in contrast, ab initio modelling should only be considered if no other options remain.}
\label{fig-3DPred-Overview}
\end{figure}

\subsection{Different classes of structure prediction methods}
 
We can classify structure prediction strategies into two categories of difficulty: template-based modelling, and template-free modelling (see \figref{3DPred-Overview}).  In the first case, it is possible to find a suitable template for the target sequence in the PDB,  as a basis for the model, whereas for template-free modelling no such experimental structure is available. Note that it may not be trivial at all to find out in which of these two categories a structure prediction problem falls. Only if we can find a close homolog -- based on sequence similarity -- in the PDB we can be sure that a template based modelling strategy will suffice; this is also referred to as homology modelling. With a template, the constraints from the alignment between the model and the template sequence, in addition to the template structure, will give sufficient constraints to build a structural model for the target sequence. 
Even in this case, small missing substructures in the alignment, e.g. loops, may require a template-free modelling strategy.

If no close homologs are available in the PDB, we may need to use more advanced template finding strategies, such as remote homology detection or fold recognition methods. 

If no suitable template is available, we will need to resort to a template-free modelling strategy. In the ``ab initio'' approach knowledge-based energy terms are used to generate structural models based  on the sequence of the template alone. Small, suitable fragments, from various PDB structures are assembled to generate possible structural models.
In some cases, we can find additional constraints, for example from experiments, such as NMR or cryoEM, or from contact prediction methods; in that case we have a much better chance of building a suitable model \cite{Moult2016b}. In fact, we could consider such constraints an alternative for the constraints provided by homology.

Lastly, several steps may be taken to refine the model, and to select the most likely model, from several model building attempts. Note that some structure prediction methods, may also include variations of model refinement and model selection steps higher up in the modelling workflow. 

\subsection{Domains}
\label{sec-domains}

So far we have implied that we may follow the above strategy for an entire protein, however, this generally is not the case. 
In fact many proteins consist of multiple domains. If this is the case, it is wise to also run one or multiple disorder prediction methods on the target sequence; Any large regions ($> 25$ residues) predicted to be disordered should be left out for further structure prediction and template finding.  

Most structure prediction methods only work well at the domain level. This means that a sequence first needs to be split in multiple domains, before we can start to make models. However, domain splitting is often ambiguous both given the sequence and the structure, while combining models built from various domains is far from trivial. In practice, this means that multiple templates might be necessary for a single target sequence and that it is difficult to resolve the orientation of the modelled domains with respect to each other.

Predicting the orientation for several domains is currently an unsolved problem, unless there is a suitable, homologous, template available -- with the domains in the same orientation.   In some cases, coarse constraints on the domain orientations such as data from small-angle scattering experiments, or distance restraints from NMR, chemical cross-links or co-evolution may help to put different homology models in the correct orientation.

Typically, it only makes sense to generate a model, be it template-based or template-free, for a single domain. In fact, in CASP model predictions are assessed per structural domain, separately. Therefore, it is essential to split the target sequence into its constituent domains -- which is a non-trivial task, particularly if no homologous templates are available for each of the domains.

\section{Assessing the quality of structure prediction methods}

Generally, as with any prediction problem, we can assess the quality of a prediction if we have a true answer to the question. Here, truth will be represented by an experimentally determined protein structure (of high quality). Fortunately, there are now (November 2017) over 120,000 deposited protein \foothref{https://www.rcsb.org/pdb/statistics/holdings.do}{structures in the PDB}. However, simply assessing how well a method performs over this set is problematic. The methods have been trained on this data set; that means that there may be a strong bias in these methods, to predict good models for sequences that are within their dataset, and therefore homologs of those. In order to truly assess a method, a completely independent data set is required. 

\subsection{Critical Assessment of protein Structure Prediction}

\begin{figure}
\includegraphics[width=\linewidth]{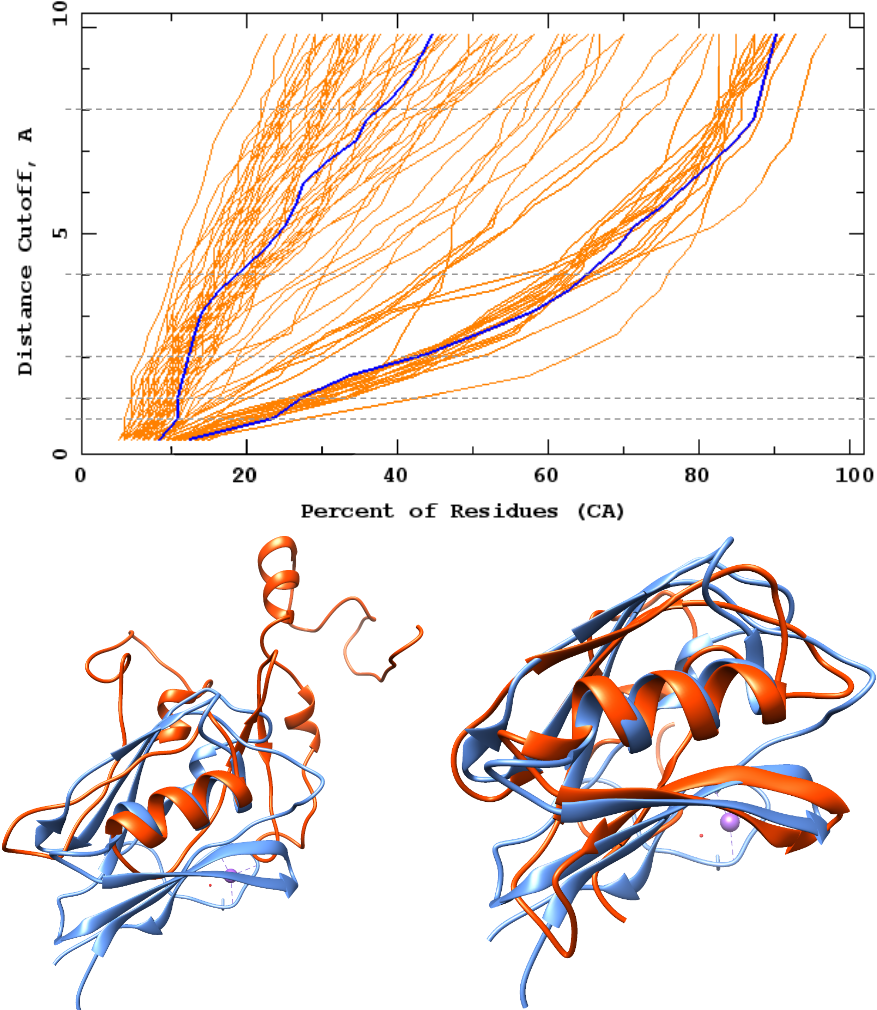}
\caption{Example of structural comparison for the target 
\href{http://predictioncenter.org/casp12/gdtplot.cgi?target=T0886-D2}{T0886-D2} and two models submitted to CASP12. 
The top panel shows individual traces for all models generated for this target; the distance cutoff (vertical axis, in \AA) is plotted against the fraction of residues (horizontal axis, in \%) that can be aligned within this cutoff. The traces were obtained from \href{http://predictioncenter.org/casp12}{predictioncenter.org/casp12}. The dotted lines indicate the thresholds used in the GDT\_TS (1, 2, 4, 8 \AA) and GDT\_HA (0.5, 1, 2, 4 \AA) scores. 
Two models are highlighted in blue: a bad model (TS236, GDT\_TS=18.90) on the left, and a good model (TS173; GDT\_TS=51.97) on the right. Both model structures are also shown in the panels below in red, superposed onto the solution crystal structure in blue 
(\href{http://www.rcsb.org/pdb/explore/explore.do?pdbId=5fhy}{PDB:5FHY}). 
Structural superposition created using LGA at 
\href{http://proteinmodel.org/AS2TS/LGA/lga.html}{proteinmodel.org/AS2TS/LGA/}~\protect\cite{Zemla2003}, 
3D visualisation using \textsl{Chimera} 1.11.2~\protect\cite{Pettersen2004}.
}
\label{fig-3DPred-GDTTS}
\remark{copyright OK -- from scratch by Anton}
\end{figure}

Every other year CASP, a Critical Assessment of Protein Structure Prediction, provides such an independent validation benchmark.
CASP is a blind test or competition: experimentalists provide sequences for which they know the structures will be solved imminently; modelling groups and servers try to predict the structure \cite{Moult1995}. Once the structure is solved, the models can be evaluated using the solution structure of the target (see also \figref{3DPred-GDTTS}). 

CASP was started because the protein structure prediction problem was claimed to have been solved several times. The problem was, that algorithms were trained on databases that contained the structures that were evaluated in benchmarking tests. CASP overcomes this problem. 

Note that the very first step in any practical structure prediction approach, should be to inspect the results from the latest CASP round \cite{Moult2016b} via the
 \foothref{http://predictioncenter.org/}{CASP website} to see what the state of the art methods are, and what their expected performance is.

\subsection{Root-Mean-Square Deviation (RMSD):}

If we want to asses the quality of a method, we need to measure the quality of the predictions made by the method. Hence, one would like to structurally compare atomic coordinates of the model and of the solution structure, and quantify the (dis)similarity.

The problem of comparing a model to to a solution structure, is less difficult than the comparison between two homologous protein structures. This is because the alignment is trivial: the model has the same sequence as the solution structure; we know which residues, and atoms should correspond in the two structures.

The easiest way to compare structures, is the calculate the Root-Mean-Square Deviation (RMSD) after a structural superpositioning 
\cite{Marti-Renom2009}. 
The superpositioning is required, because two arbitrary structures will typically not be positioned at coordinates suitable for comparison; first a translation and rotation needs to be applied to one of the two structures, to minimise the RMSD; the resulting RMSD after superpositioning can be used as a dissimilarity measure.

The Root-Mean-Square Deviation (RMSD) calculates the squared difference between two sets of atoms, and can be defined as follows:

\begin{align*}
RMSD(v,w) & = \sqrt{\frac{1}{n}\sum_{i=1}^n \left \| v_i-w_i \right \|^2} \\ 
		  & = \sqrt{\frac{1}{n}\sum_{i=1}^n 
          		\left ( v_{ix}-w_{ix} \right )^2 + 
                \left ( v_{iz}-w_{iz} \right )^2 + 
                \left ( v_{iz}-w_{iz} \right )^2}
\end{align*}

Here, $v_i$ is the position vector of $i^{th}$ atom of structure $v$;
$w_i$ is the position vector of $i^{th}$ atom of structure $w$; and 
$n$ is the total number of aligned atoms.

The RMSD takes the average over all aligned pairs. In protein structures typically one representative atom per residue is chosen, such as C$\alpha$ or C$\beta$.

\subsection{GDT -- Global Distance Test}

If a model gets a loop very wrong, it tends to stick out and can be positioned very distant from the true structure, even though the remaining structure may be reasonably accurate. This partial outlier weighs heavily on the average distance calculated. Hence the RMSD is over sensitive to such outliers.

The global distance test total score (GDT\_TS) is a more robust structural similarity measure that is well defined given an alignment between two structures. The key idea is to count the number of residues that can maximally be fitted within a certain distance cutoff, see also \figref{3DPred-GDTTS}. The GDT score will therefore produce a percentage. In the formula below, the final score is the average over four different distance cutoffs (1, 2, 4, 8 \AA).

\begin{equation}
	GDT\_TS = \frac14 \sum_{v=1,2,4,8\text{\AA}}\frac{G(v)}{t}
\end{equation}

Here, $G(v)$ is the number of aligned residues within given RMSD cutoff $v$ (in {\AA}ngstrom -- $10^{-10}m$) and $t$ is the total number of aligned residues. 
A related score called GDT\_HA was introduced in CASP some time ago \cite{Read2007} using stricter distance cutoffs (0.5, 1, 2, 4 \AA), to cater for targets in the template-based modelling category where very high accuracies can be realized:

For a typical ``difficult'' CASP target no model even comes close to the experimentally solved structure; typical results would be similar to the left-most model in \figref{3DPred-GDTTS}. If we have a look at the latest CASP results one will see that a performance of GDT\_TS$<20\%$ is not an exception. In other words, the protein structure prediction problem has NOT yet been solved, especially not if one considers targets without a good template structure.

\begin{figure}
\includegraphics[width=\linewidth]{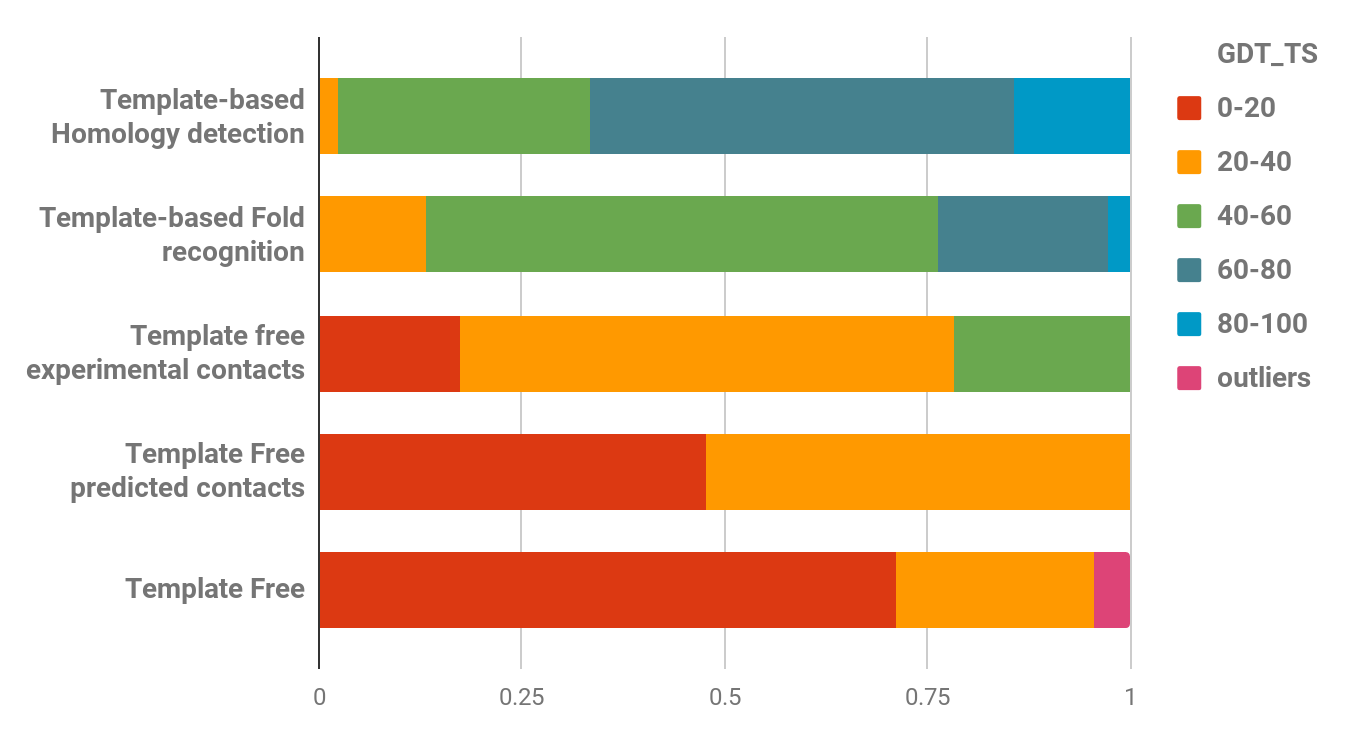}
\caption{Distribution of GDT\_TS scores for the different model categories in CASP11 for template-based \cite{Modi2016a}, template-free with contact information \cite{Kinch2016b} and template-free \cite{Kinch2016a}. The legend coloring corresponds to the GDT\_TS scores, the bars indicate the fraction of models in each GDT\_TS range for the six categories (GDT\_TS scores for \cite{Modi2016a} were estimated from the reported GDT\_HA scores using their Figure 4A). ``Outliers'' targets have unusually high GDT\_TS due to being very short ($\sim 50$ residue) with extended structures. Targets selected for server prediction (top bar) were considered easier than those for human prediction (second from top), average sequence identity was 26\% vs. 20\%, respectively. It is clear that overall prediction accuracy sharply declines going down this list of categories. For template-free modelling, the quality of contact information used is crucial. Experimental information (from chemical cross linking or simulated NMR) can give reasonable models. Predicted contacts do not guarantee that an acceptable model can be obtained, but without even predicted contacts, more than two-thirds of models are at most 20\% correct.}
\label{fig-3DPred-GDTstats}
\remark{copyright OK -- from scratch by Anton}
\end{figure}

\subsection{How difficult is it to predict?}

Overall, if one can find a good template, the quality of the predicted model will be relatively good. CASP results show that for homology modelling based on close homologues, it is possible to obtain models similar to the experimentally determined structure \cite{Moult2016b}. The modelled structure will typically have a good accuracy for the regions that can be well aligned between the target and template (using the sequences). The top two bars in \figref{3DPred-GDTstats} shows that one may expect the majority of such models to be accurate for $>50\%$ of their residues. Gaps in an alignment will typically lie in loop regions of a structure and are more difficult to model. So, if we are interested in a large loop region that is not present in our template, we still may not be able to answer our scientific question with the resulting model structure \cite{Moult2016b}.

If no acceptable template can be found, the chances of successfully answering our scientific question will become very low. As a last resort, ab initio modelling can provide us with structural models. Typically, ab initio methods use very small templates from various proteins (see \figref{3DPred-Overview}). The state of the art is that on average one may expect to find one structure that looks somewhat like the solution structure for the target among the top five or ten models \cite{Moult2016b}. However, be aware that the best model is typically not recognised as being the best through the scores of the prediction program. In \figref{3DPred-GDTstats} one sees that very clearly in the bottom few bars: without template, even with predicted contacts, one may have less than 20\% of the structure correct in the majority of models; even in the best cases at most 40\% of the residues are modelled accurately.

\subsection{For which gene sequences can we predict a three-dimensional structure?}

If and only if there is a structure of a homologous protein present in the PDB, it is possible to generate a structural model of reasonable accuracy. Based on this notion, we can estimate for which (fraction of) gene sequences it is possible to predict a structure. This way it has been estimated that for a about 44\% of residues in Eukaryotic gene sequences, we cannot yet make a homology model, and 15\% of these residues lie within a gene for which we can not make a homology model for a single domain \cite{Perdigao2015}. Especially membrane proteins are underrepresented in the PDB, due to the experimental difficulty of determining these structures. Note that these residues, may also lie in natively disordered regions (see also \secref{single-native-fold}).

Similarly, it is possible to predict the range of protein structures present in an organism, based on the gene sequences their completed genome. This reveals that there is a subset of protein structures, that is present in nearly all organisms, for example TIM-barrels or Rossmann-folds \cite{Abeln2005,Edwards2013}. Nevertheless, there is also a group of structures that is extremely lineage specific. It is to be expected that for this type of protein structures, many new structures remain to be discovered. This also implies that it will remain difficult to find suitable templates for homology modelling for these lineage specific protein families.

\subsection{How accurate do we need to be?}
We already mentioned that we may approach the modelling of a protein structure of interest differently, depending on the biological question we want to ask,e.g. which residues are likely to be crucial for the functioning of the protein. 
Sometimes an answer to the research question may be possible in a simpler way, without full-scale prediction of the protein structure, e.g. by direct prediction of the impact of certain mutations or of protein-protein interaction sites. Examples of fully-automated webservers that do just that, are HOPE --  \cite{Venselaar2010} and SeRenDIP  \cite{Hou2017}. In some cases, a rough homology model inspires the understanding of experimental results, spurring forward the project and eventually ending with crystal structures highlighting the protein function (in this case, protein-protein interactions) of interest \cite[e.g.,][]{DeVries-vanLeeuwen2013}. Also, specifically for enzymes, such as for example cytochromes P450, modelling of the protein structure should be done combination with that of the ligand \cite{Graaf2005}.

In CASP11, three functional aspects were explictly scored, selected on being able to qualitatively evaluate them: multimeric state, (small) ligand binding, and mutation impact. Targets were selected that in solved crystal structure were dimeric, or had a ligand bound, or where from the crystallographers or in literature interest was expressed for evaluating mutants \cite{Huwe2016}.

For prediction of dimer structures, only in two cases out of ten a dimer model with reasonable accuracy could be generated for the majority of monomer model structures \cite{Huwe2016}. In the critical assessment of prediction of protein interaction (CAPRI) between 30-80\% of models were of `acceptable' or `medium' quality for easy dimer targets, while for harder targets (difficult dimers, multimers and heteromers), this fraction dropped to below 10\% \cite{Lensink2016}. Encouragingly, it was seen that also structure models of lower quality could sometimes lead to acceptable or even medium quality models of the bound proteins \cite{Lensink2016}.

For ligand binding, it was found that the accuracy of even the best models ($\sim 2$\AA) are not good enough for accurate ligand docking; the best ligands were around $5${\AA} RMSD \cite{Huwe2016}. Something similar was found for mutation impact prediction; for most targets, model accuracy did not correlate with accuracy of impact prediction \cite{Huwe2016}. Apparently, either homology models are not yet accurate enough for this purpose, or methods are tuned to particular characteristics of crystal structures. 

\section{Is there such a concept as a single native fold?}
\label{sec-single-native-fold}
Before we conclude, we should consider a more physical description of protein structure. In fact, protein folding from a physical point of view is a very interesting process: given a sequence, a protein tends to fold always, and exactly into the same functional structure. In material design, it is extremely difficult to mimic such high specificity.
The apparent observation of folding specificity also leads to the question, is there such a concept as a single native fold? Or, more pragmatically, is sequence-to-structure truly a one-to-one relation? 

In fact, if one wants to start making quantitative predictions, such as the stability of a protein fold, or the binding strength between two proteins in terms of free energy, it is much more helpful to think in ensembles of structural configurations for a protein sequence \cite[e.g.][]{May2014,Pucci2017}. The probability to find a protein in a specific ensemble of structural configurations will depend on conditions such as the presence or absence of binding partners, the pressure, the pH or the temperature \cite[e.g.][]{vanDijk2015,vanDijk2016}. There are a few specific cases,  common cases, for which even the functional or biologically relevant structural ensembles do not resemble a single globular folded structure.

\subsection{Disordered proteins}
Not all proteins fold into single configurations, some proteins stay natively unfolded, i.e. they can take up a a large variety of more extended, and very different configurations \cite{Uversky2000,Meszaros2007}.  Some disordered regions contain elements that do form stable structures upon binding. The regions that remain disordered are thought to be important to prevent aggregation within the cell \cite{Abeln2008}.
Missing residues in X-ray structures are typically removed for crystallization; for this reason disorder prediction methods have been developed. Disordered regions are relatively easy to predict in protein sequences just like secondary structures; broadly speaking, prediction can be based on the large amount of charged/polar (hydrophilic) amino acids in combination with the presence of amino acids that disrupt the secondary structure (proline and glycine) in these regions \cite{Oldfield2005,Wang2016}. We know sequences of many proteins contain large disordered segments (33\% of eukaryotic, 2\% archaeal, and 4\% bacterial proteins).


\subsection{Allostery and functional structural ensembles}

It is important to realize that one protein, typically, does not correspond to one defined three-dimensional structure. Disordered regions or proteins are one particularly salient case, but also proteins which fold into specific three-dimensional configurations, may exist in multiple functional states each with a specific structure. The biological question of interest dictates which state is relevant. Most proteins have only been crystallized in one particular state, and often it is not known to which biological condition this crystal structure may correspond. One may have cases where a homology model of the relevant state may be preferred over a crystal structure of a different or unknown state \cite[e.g.,][]{Graaf2005}.

\subsection{Amyloid fibrils}
Lastly, we should consider a competing state of folded proteins: the aggregated state, where multiple peptide chains clog together in fibrillar structures or amorphous aggregates. Amyloid fibres are formed by $\beta$-strands formed between different protein or peptide (small protein) chains. Fibril formation is associated with various neurodegenerative diseases, such as Alzheimer's, Creutzfeldt-Jakob and Parkinson's \cite{Chiti2006}. In fact, the fibrillar state is more favorable than the state of separately folded structures for several protein types. The general cellular toxicity of such aggregates, puts evolutionary pressure on avoiding structural characteristics on the surface of proteins; hence it is extremely rare to observe solvent accessible $\beta$-strand edges or large hydrophobic surface patches \citep{Richardson2002,Abeln2011}. The propensity proteins have to form Amyloid fibrils is relatively easy to predict \cite{Grana-Montes2017}. However, reference databases are still small so it is difficult to verify such methods. 

\section{Acknowledgements}
We thank Nicola Bonzanni, Kamil K.\@ Belau, Ashley Gallagher, Jochem Bijlard for insightful discussions and critical proofreading of early versions.

\bibliographystyle{apalike}
{\small\raggedright
\bibliography{prot3dpred}
}

\printindex

\end{document}